\documentstyle[aps,twocolumn,epsf,psfig]{revtex}

\begin{document}
\draft

\title {Electron-Phonon Correlations, Polaron Size, \\
 and the Nature of the Self-Trapping Transition}

\author{A.~H.~Romero${^{0,1,2}}$
David W. Brown${^3}$ and Katja Lindenberg${^{1,3}}$}

\address
{${^1}$
Department of Chemistry and Biochemistry,\\
University of California, San Diego, La Jolla, CA 92093-0340} 

\address
{${^2}$
Department of Physics,\\
University of California, San Diego, La Jolla, CA 92093-0354}

\address
{${^3}$
Institute for Nonlinear Science,\\
University of California, San Diego, La Jolla, CA 92093-0402}

\date{\today} 

\maketitle

\begin{abstract}

We analyze electron-phonon correlation functions measured in 1D polaron ground states of the Holstein Hamiltonian using the Global-Local variational method.
The spatial collapse of electron-phonon correlations is found to occur in concert with transition behavior in other polaron properties, providing mutually confirming evidence for a self-trapping line in 1D.
The spatial extent of electron-phonon correlations is used to quantify polaron size, and is analyzed over a wide range of parameters.
Distinct scaling behaviors are found to be characteristic of the region below the self-trapping transition and above it, contrary to some widely-held expectations and leading naturally to the notion of the polaron size as an order parameter for a self-trapping transition that becomes critical in the adiabatic limit.

\end{abstract}

\pacs{PACS numbers: 71.38.+i, 71.15.-m, 71.35.Aa, 72.90.+y}

\narrowtext

\footnotetext[0]{Present address: Max-Planck Institut f\"{u}r Fest\-k\"{o}rper\-forschung, Heisenbergstr. 1, 70569 Stuttgart, Germany}

Polarons are the quantum quasiparticles describing the states of excitations correlated with the deformation or polarization quanta of a host medium.
Perhaps the single most intuitively-accessible polaron property is its size; however, the size of the polaron is one of its properties that is most elusive.
Generally, we do not measure it directly, but typically impute to it some qualitative character (e.g., ``large'' or ``small'') by interpreting measurements of more directly-accessible properties within some theoretical view of the problem, often reduced to rules of thumb.

In this paper, rather than rely upon rules of thumb and qualitative characterization, we subject the notion of polaron size to direct quantitative test.
A large volume of numerical data is analyzed to put numerical value to measures of the polaron width over a broad range of the polaron phase diagram, and the essential characteristics of polaron size are mapped out.
The volume, scope, and accuracy of our data makes it possible to discern trends that indicate new ``rules of thumb'' that challenge existing thinking about polaron size.

The size of the polaron is not uniquely-definable owing both to the inherent complexity of the polaron quasiparticle and to the strictly delocalized character of polaron eigenstates in translationally-invariant lattices.
There is nonetheless no shortage of opinion regarding the spatial structure of the polaron, much of which is tied to the adiabatic approximation \cite{Rashba,Holstein59a,Toyozawa,Holstein,Kabanov93,Song96}; some of this is based on localized-states, the balance on delocalized states built up from localized states.

We work in a strictly delocalized-state picture, wherein we construct polaron Bloch states as superpositions of localized functions (``form factors'') that are the vehicles through which local electron-phonon correlation structure is conveyed into the delocalized momentum eigenstate.
The quality of result is directly attributable to the appropriateness of the local form factor.
Much work has been done on polaron states as localized states, from which point of view the form factors we use are themselves viable localized polaron states.
While some similarity of result between such approaches can be expected, it is essential to the following that such similarity is limited.
When integrated into the delocalized Bloch state, the ``identity'' of the form factor is lost and one can no longer characterize the properties of the quantum state in terms uniquely associated with it.
Local properties can be probed, however, with correlation functions that reveal the internal structure implicit in the delocalized state; thus, we characterize polaron size through the use of electron-phonon correlation functions that reveal the structure of the ``phonon cloud'' or co-moving lattice distortion that surrounds the electron at every instant.

We use the Global-Local variational method to determine a large number of such ground states over broad regions of the polaron phase diagram.
The particulars of the Global-Local method and other results obtained through its use have been detailed elsewhere \cite{Brown97b,Romero98g,Romero98a,Romero98c,Romero98d,Romero98e,Romero98f,Romero99}.
The Global-Local method has been demonstrated to be consistent with both weak-coupling and stong-coupling perturbation theories in the appropriate regimes, and over the regimes studied here, to be in broad quantitative agreement with other high-quality methods such as
cluster diagonalization ~\cite{Capone97,Wellein97a,deMello97,Alexandrov94a},
density matrix renormalization group ~\cite{Jeckelmann98a}, and
Monte Carlo simulations ~\cite{Lagendijk,Kornilovitch98a}.

We limit ourselves to 1D and use the traditional Holstein Hamiltonian~\cite{Holstein59a} defined by

\begin{eqnarray}
\hat{H} &=& - J \sum_n a_n^{\dagger} ( a_{n+1} + a_{n-1} ) + \hbar \omega \sum_n b_n^{\dagger} b_n \nonumber \\
&&- g \hbar \omega \sum_n a_n^{\dagger} a_n ( b_n^{\dagger} + b_n ) ~,
\end{eqnarray}
in which $a_n^\dagger$ creates a single electronic excitation in the rigid-lattice Wannier state at site $n$, and $b_n^\dagger$ creates a quantum of vibrational energy  $\hbar \omega$ in the Einstein oscillator at site $n$. 
The hopping matrix element connecting nearest neighbors is given by $J$, and $g$ is the electron-phonon coupling  strength.
We will use $\hbar \omega$ as the unit of energy, in terms of which the two independent model parameters are defined (hopping integral $J/\hbar \omega$ and electron-phonon coupling $g$).

\begin{figure}[!hb]
\begin{center}
\leavevmode
\epsfxsize = 3.6in
\epsffile{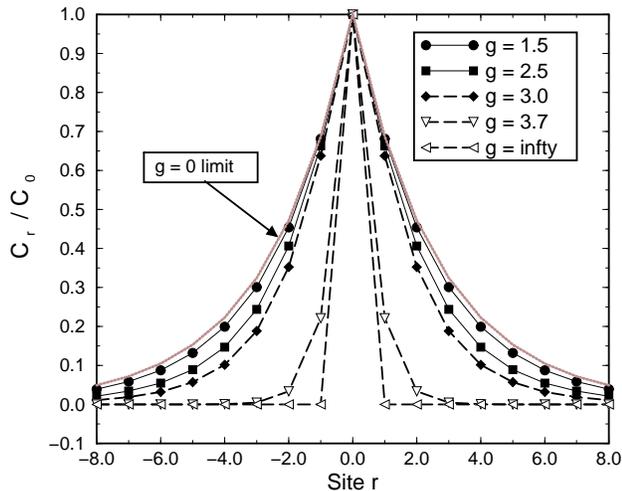}
\end{center}
\caption
{
Electron-phonon correlation function $C_r$ as determined by the Global-Local method.
$J / \hbar \omega = 7.0$ for assorted values of $g$.
Open symbols lie in the small polaron region, solid symbols in the large polaron region.
}
\label{fig:correljs}
\end{figure}

The substance of our results should not depend greatly on precisely what correlation function we use, provided that electron-phonon relationships are spatially resolved.
We use the function
\begin{equation}
{C}_r \equiv \frac 1 {2g} \sum_n \langle a^{\dagger}_n a_n ( b^\dagger_{n+r} + b_{n+r} ) \rangle ~,
\end{equation}
normalized such that $\sum_r C_r = 1$.
Examples of $C_r$ at general $J/\hbar\omega$ and $g$ obtained by the Global-Local variational method are shown in Figure~\ref{fig:correljs} \cite{Romero98g,Romero99}.
These correlation functions are strongly exponential over four decades of amplitude in the site index $r$ except in the intermediate coupling region where mild deviations from exponential decay are found.

\begin{figure}[htb]
\begin{center}
\leavevmode
\epsfxsize = 3.6in
\epsffile{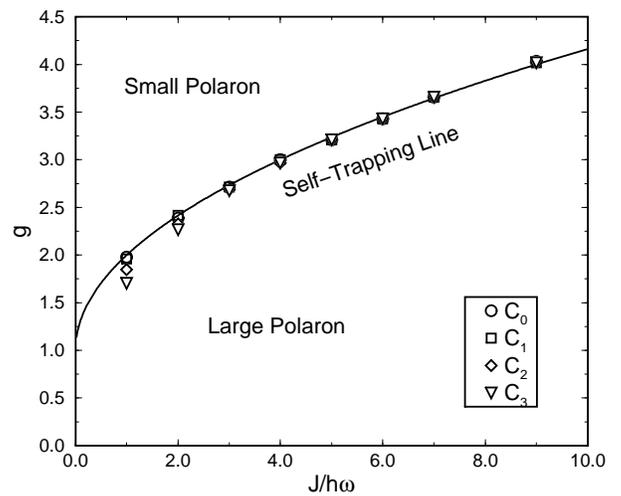}
\end{center}
\caption
{
Polaron phase diagram, showing the location of the self trapping line $g_{ST}$ and the specific loci due to Global-Local measurements of $C_{r = 0,1,2,3}$.
}
\label{fig:phasec0123}
\end{figure}

The dependence of individual $C_r$ on model parameters can be used to determine the location of the self-trapping transition.
Through a curvature analysis of a large volume of Global-Local data, we have identified the self-trapping transition with the point of most rapid {\it decrease} in $C_r$ w.r.t. $g$ at fixed $J/\hbar\omega$ for several sites near the center of the polaron.
Figure~\ref{fig:phasec0123} shows that these loci for $C_{r = 0,1,2,3}$ cluster tightly about the same self-trapping line ($g_{ST}$) as has been previously identified through numerous other polaron properties \cite{Romero98g,Romero98c,Romero98e,Romero99}, assuring that all these analyses are identifying the same fundamental event.
The apparent self-trapping trends in all these are well located by the empirical curve $g_{ST} = 1+\sqrt{J/\hbar\omega }$ as shown in Figure~\ref{fig:phasec0123}.

The self-trapping line $g_{ST}$ is a very important feature in all these considerations because it sets the coupling scale around which the most physically-significant features are organized.
At large $g$, above the self-trapping transition, the polaron is extremely compact up to the self-trapping line where a fairly rapid spreading of correlations commences; however, the spreading of the correlations saturates at a finite width in the weak-coupling limit.
This saturation phenomenon proves to be a crucial finding at variance with many widely-held views about the nature of large polarons \cite{Romero98g,Romero98d,Romero98f,Romero99}.

Our basic tool for quantifying polaron size is the variance of the correlation function
\begin{equation}
\sigma^2 = \sum_r r^2 C_r ~.
\end{equation}

An important first result from weak-coupling perturbation theory \cite{Romero98g,Romero98d,Romero98f,Romero99} is
\begin{equation}
\lim_{g \rightarrow 0} \sigma^2 = {\sigma_0^2} = \frac {2J} {\hbar \omega} ~.
\label{eq:sigmawc}
\end{equation}
This limiting behavior is what is seen in Figure~\ref{fig:correljs}, where the spreading of electron-phonon correlations saturates at a finite width as coupling vanishes.

On the other hand, above the self-trapping transition it can be shown (for example, using strong-coupling perturbation theory \cite{Romero98g,Romero99}) that
\begin{equation}
\lim_{J/\hbar\omega \rightarrow \infty} \frac {\sigma^2} {\sigma_0^2} = 0 ~~~~~~~~~~ g > g_{ST}
\end{equation}
This result involves the fact that being ``above'' the self-trapping transition while $J/\hbar\omega$ diverges requires that $g$ diverge as well in a sufficiently rapid manner.

With these properties in mind, we scale our variance data so that they can be meaningfully organized relative to the self-trapping transtion.
The weak-coupling variance $\sigma_0^2$ appears to be the natural quantity with which to scale variances, and $g_{ST}$ the natural quantity with which to scale the electron-phonon coupling.

\begin{figure}[htb]
\begin{center}
\leavevmode
\epsfxsize = 3.6in
\epsffile{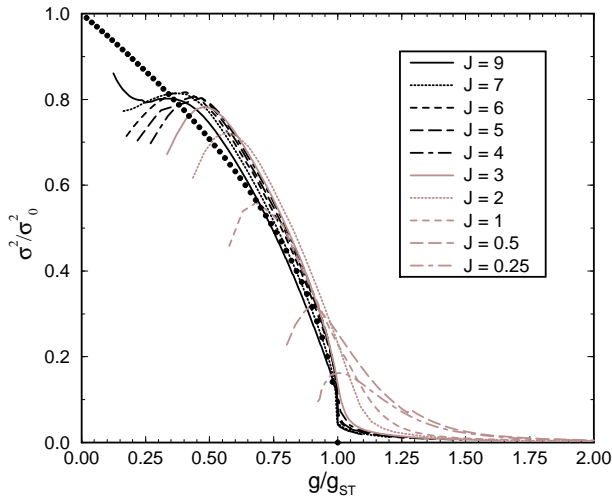}
\end{center}
\caption
{
Scaled correlation variance $\sigma^2 / \sigma_0^2$ as results from the Global-Local variation.
Variance data are truncated as discussed in the text.
}
\label{fig:width}
\end{figure}

Using the Global-Local variational method, we have computed $\sigma^2$ for approximately 1200 polaron ground states and have compiled these results in Figure~\ref{fig:width} using the above scaling conventions.

These data deteriorate at the weak coupling end of each curve where the Global-Local method encounters systematic difficulties.
The variance as a measure of spatial structure {\it amplifies} contributions from tails of the polaron wave function, indiscriminately amplifying any errors lurking in that part of the numerical solution.
This poses problems for variational methods, since the variational energy is least sensitive to the detail of such tails, with the result that relative errors are largest in this part of the solutions.
Similarly, any finite-size effects present are most deleterious in the wave function tails.
For such reasons, only the {\it descending} part of each numerically-determined curve is to be regarded as reliable; we have left a small ascending arc at the weak-coupling end of each curve only to render clearly where and how the deterioration of results sets in.

Focussing therefore on the {\it descending} data in Figure~\ref{fig:width}, we are led to distinguish two apparent facts:
1) that the variances above the self-trapping transition ($g> g_{ST}$) appear to be converging toward zero (consistent with SCPT), and
2) that the variances below the self-trapping transition ($g<g_{ST}$) appear to be converging toward finite values (consistent with WCPT).
These findings indicate a difference in {\it scaling properties} above and below the transition that suggests the emergence of critical behavior in the adiabatic limit.
Relative to such emergent critical behavior, the limiting variance curve below the transition has the characteristics of an {\it order parameter}.
What is most significant in this is not the detailed form of that limiting curve, which we do not suggest our data is sufficiently complete or accurate to determine, but merely that such a limiting curve appears to {\it exist}.

A fair characterization of the apparent trends in convergence is contained in the function
\begin{eqnarray}
\frac {\sigma^2} {\sigma_0^2} \rightarrow {\cal{F}} \left( \frac g {g_{ST}} \right) &=& \sqrt{1 - \frac g {g_{ST}}} ~~~~~ g < g_{ST} \nonumber \\
&& \nonumber \\
&=& ~~~ 0 ~~~~~~~~~~ ~~~~~ g> g_{ST}
\label{eq:orderbelow}
\end{eqnarray}
as indicated by the beaded guide curve in Figure~\ref{fig:width} and under a different scaling convention (see below) in Figure~\ref{fig:orderpar}.
\begin{figure}[htb]
\begin{center}
\leavevmode
\epsfxsize = 3.6in
\epsffile{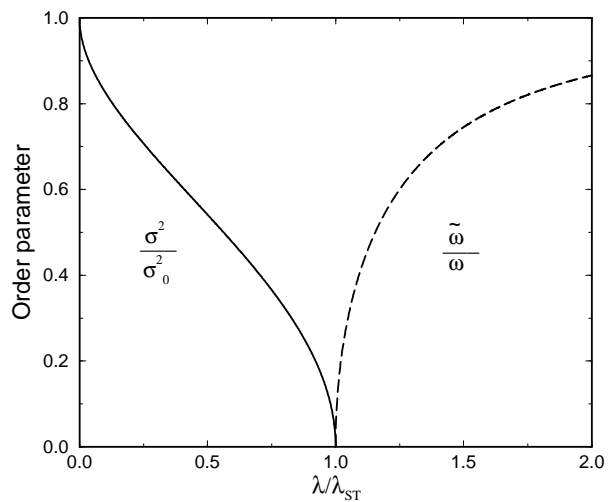}
\end{center}
\caption
{
Order parameters characterizing the adiabatic limit.
Solid curve, Eqn. \ref{eq:orderbelow}.
Dashed curve, Eqn. \ref{eq:orderabove}.
}
\label{fig:orderpar}
\end{figure} 
Thus, we find that the polaron size below the self-trapping transition is roughly proportional to that found in the weak coupling limit, with the constant of proportionality dependent primarly on the ``distance'' from the self-trapping transition.

This finding is consistent with and complementary to a well-known result of adiabatic strong-coupling theory \cite{Alexandrov95} that identifies a soft vibrational mode whose existence is synonymous with the existence of the self-trapped state.
The soft mode frequency $\tilde{\omega}$ is interpretable as an order parameter for the self-trapping transition.
In terms of the composite parameter $\lambda = g^2 \hbar \omega / 2J$ common in strong-coupling theory,
\begin{eqnarray}
\frac {\tilde{\omega}} {\omega} &=& ~~~ 0 ~~~~~~~~~~ ~~~~~ \lambda < \lambda_{ST} \nonumber \\
&& \nonumber \\
&=& \sqrt{1 - \frac {\lambda_{ST}^2} {\lambda^2} } ~~~~~ \lambda > \lambda_{ST}
\label{eq:orderabove}
\end{eqnarray}
with $\lambda_{ST} = 1/2$ in 1D.
(See Figure~\ref{fig:orderpar}.)

It is perhaps not surprising that the polaron size should have the characteristics of an order parameter, since the essential physical notion underlying the self-trapping transition is that of spatial collapse; however, according to some widely-held and long-established views, such collapse is not expected to occur in 1D.
Under the adiabatic approximation in 1D, the size-scaling behavior that we have here found to be limited to the region {\it above} the transition is found instead at {\it all} coupling strengths, without undergoing any transmutation into a distinct weak-coupling behavior.
Consequently, the ``rules of thumb'' given by the adiabatic approximation to characterize the size of the large polaron in 1D disagree with the ``rule of thumb'' that can be taken from Eqn.~\ref{eq:orderbelow}.

The discrepancies between the traditional notion of the large polaron and what we find here are pervasive and, in regimes, severe.
Necessarily, we are led to supplant the traditional notion of the large polaron as a broad ``soft'' state that responds rather strongly to changes in electron-phonon coupling strength with that of a less broad, ``stiff'' state that is relatively insensitive to changes in electron-phonon coupling strength.

Although the picture of the self-trapping transition that emerges from this work has so far been demonstrated only in 1D, many of the elements and relationships contributing to it are already known to hold essentially unchanged in higher dimensions \cite{Romero98d,Romero98f,Lagendijk,Kornilovitch98a,Romero99a}.
We expect $\sigma^2 / \sigma_0^2$ to continue to constitute an order parameter in the adiabatic limit and for the order parameter to provide a reasonable ``rule of thumb'' for the correct size scaling of the large polaron in each dimension.

\section*{Acknowledgement}

This work was supported in part by the U.S. Department of Energy under Grant No. DE-FG03-86ER13606.

\end{document}